\documentstyle[psfig,graphicx,color]{mn}

\newif\ifAMStwofonts

\def\simlt{\lower.5ex\hbox{$\; \buildrel < \over \sim \;$}}
\def\simgt{\lower.5ex\hbox{$\; \buildrel > \over \sim \;$}}

\ifoldfss
  \ifCUPmtlplainloaded \else
    \NewTextAlphabet{textbfit} {cmbxti10} {}
    \NewTextAlphabet{textbfss} {cmssbx10} {}
    \NewMathAlphabet{mathbfit} {cmbxti10} {} 
    \NewMathAlphabet{mathbfss} {cmssbx10} {} 
  \fi
  \ifAMStwofonts
    \ifCUPmtlplainloaded \else
      \NewSymbolFont{upmath} {eurm10}
      \NewSymbolFont{AMSa} {msam10}
      \NewMathSymbol{\upi}     {0}{upmath}{19}
      \NewMathSymbol{\umu}     {0}{upmath}{16}
      \NewMathSymbol{\upartial}{0}{upmath}{40}
      \NewMathSymbol{\leqslant}{3}{AMSa}{36}
      \NewMathSymbol{\geqslant}{3}{AMSa}{3E}

    \fi
  \fi
\fi 

\ifnfssone
  \newmathalphabet{\mathit}
  \addtoversion{normal}{\mathit}{cmr}{m}{it}
  \addtoversion{bold}{\mathit}{cmr}{bx}{it}
  \newmathalphabet{\mathbfit} 
  \addtoversion{normal}{\mathbfit}{cmr}{bx}{it}
  \addtoversion{bold}{\mathbfit}{cmr}{bx}{it}
  \newmathalphabet{\mathbfss} 
  \addtoversion{normal}{\mathbfss}{cmss}{bx}{n}
  \addtoversion{bold}{\mathbfss}{cmss}{bx}{n}
  \ifAMStwofonts
    \ifCUPmtlplainloaded \else
      \UseAMStwoboldmath
      \makeatletter
      \new@mathgroup\upmath@group
      \define@mathgroup\mv@normal\upmath@group{eur}{m}{n}
      \define@mathgroup\mv@bold\upmath@group{eur}{b}{n}
      \edef\UPM{\hexnumber\upmath@group}
      \new@mathgroup\amsa@group
      \define@mathgroup\mv@normal\amsa@group{msa}{m}{n}
      \define@mathgroup\mv@bold\amsa@group{msa}{m}{n}
      \edef\AMSa{\hexnumber\amsa@group}
      \makeatother
      \mathchardef\upi="0\UPM19
      \mathchardef\umu="0\UPM16
      \mathchardef\upartial="0\UPM40
      \mathchardef\leqslant="3\AMSa36
      \mathchardef\geqslant="3\AMSa3E
    \fi
  \fi
\fi 

\ifnfsstwo
  \DeclareMathAlphabet{\mathbfit}{OT1}{cmr}{bx}{it}
  \SetMathAlphabet\mathbfit{bold}{OT1}{cmr}{bx}{it}
  \DeclareMathAlphabet{\mathbfss}{OT1}{cmss}{bx}{n}
  \SetMathAlphabet\mathbfss{bold}{OT1}{cmss}{bx}{n}
  \ifAMStwofonts
    \ifCUPmtlplainloaded \else
      \DeclareSymbolFont{UPM}{U}{eur}{m}{n}
      \SetSymbolFont{UPM}{bold}{U}{eur}{b}{n}
      \DeclareSymbolFont{AMSa}{U}{msa}{m}{n}
      \DeclareMathSymbol{\upi}{0}{UPM}{"19}
      \DeclareMathSymbol{\umu}{0}{UPM}{"16}
      \DeclareMathSymbol{\upartial}{0}{UPM}{"40}
      \DeclareMathSymbol{\leqslant}{3}{AMSa}{"36}
      \DeclareMathSymbol{\geqslant}{3}{AMSa}{"3E}
    \fi
  \fi
\fi 

\ifCUPmtlplainloaded \else
  \ifAMStwofonts \else 
    \def\upi{\pi}
    \def\umu{\mu}
    \def\upartial{\partial}
  \fi
\fi

\title[Spectral mapping of IP~Pegasi]
		{Spectral mapping of the spiral structures in IP~Pegasi on the 
		decline from an outburst}
\author[R. Baptista et~al.]
       {Raymundo Baptista$^1$, Carole A.\ Haswell$^2$ and Gino Thomas$^3$ \\
       $^1$ Dept.\ de F\'\i sica, Universidade Federal de Santa Catarina,
       Trindade, 88040-900, Florian\'opolis - SC, Brazil, 
       email: bap@fsc.ufsc.br \\
       $^2$ Dept.\ of Physics and Astronomy, The Open University, Walton
       Hall, Milton Keyes, MK7 6AA, UK, email: c.a.haswell@open.ac.uk \\
       $^3$ St.\ John's College, 1160 Camino Cruz Blanca, Santa Fe,
       NM 87505, USA, email: grt@mail.sjcsf.edu \\ }

\date{Accepted 2002 March 21; Received 2002 March 5; in original form 2001 October 31}

\pagerange{1--12}
\pubyear{2002}

\begin{document}
\label{firstpage}

\maketitle

\begin{abstract}

We report eclipse mapping of time resolved spectroscopy of the dwarf 
nova IP~Pegasi on the late decline from the May 1993 outburst.
The continuum light curves exhibit an asymmetric `V' shape with broad 
bulges and results in eclipse maps with two asymmetric arcs extended 
both in radius [$R\simeq (0.2-0.6)\; R_{L1}$, where $R_{L1}$ is the
distance from the disc centre to the inner lagrangian point] and in 
azimuth (by $\simeq 90\degr$), interpreted as a two-armed spiral shock.
The spiral arms are thus still visible some 8 days after the onset
of the outburst. Their fractional contribution to the continuum 
emission, 12 per cent of the total light, is similar to that measured 
close to outburst maximum, whereas their orientation is rotated by 
$58\degr$ with respect to the spirals seen in the eclipse map at 
outburst maximum.
The radial temperature distribution computed from the spiral-free disc 
regions is flat, with temperatures of about $5000\,K$ at all disc radii.

Velocity-resolved light curves across the H$\alpha$ and the He\,I 
lines show the classical rotational disturbance, with the blue side of 
the line being eclipsed earlier than the red side. The differences
between the H$\alpha$ and the He\,I maps are significant.
The spiral arms are clearly seen in the He\,I maps, with the receding 
arm being stronger in the red side while the approaching arm is stronger 
in the blue side of the line.
The analysis of the H$\alpha$ maps suggests that this emission arises 
mainly from a large and vertically-extended region which we interpret as
an outflowing (and spiraling) disc wind.
The H$\alpha$ emission-line surface brightness is flat in the inner 
disc regions ($I_\nu \propto R^{-0.3}$ for $R<0.3\;R_{L1}$) but decreases 
sharply with radius in the outer disc ($I_\nu \propto R^{-2}$ for $R>0.3\; 
R_{L1}$).
The spectrum of the uneclipsed light is dominated by a strong, 
blueshifted and narrow H$\alpha$ emission line superimposed on a red 
continuum and can be understood as a combination of emission from an
M5V secondary star plus optically thin emission from the outer parts
of the vertically-extended disc wind.
The inner disc regions show an emission line spectrum with a strong 
[$EW= (100\pm 2)$~\AA] and broad ($FWZI\simeq 3000\; km\;s^{-1}$) 
H$\alpha$ component superimposed on a flat continuum.
This is in marked contrast with the results from the spectral mapping of
nova-like variables of comparable inclination and mass ratio and suggests 
that intrinsically different physical conditions hold in the inner disc 
regions of outbursting dwarf novae and nova-like systems.

\end{abstract}


\section{Introduction}

Many aspects of accretion disc physics are best studied in 
mass-exchanging binaries such as non-magnetic Cataclysmic
Variables (CVs). In these close binaries mass is fed to a white dwarf
(the primary) by a Roche lobe filling companion star (the secondary) 
via an accretion disc, which usually dominates the ultraviolet and 
optical light of the system (Frank, King \& Raine 1992; Warner 1995). 

Accretion discs in CVs cover a range of accretion rates, \.{M}, 
and different viscosity regimes.
For example, the sub-class of dwarf novae comprises low-mass transfer 
CVs showing recurrent outbursts (of 2--5 magnitudes, on timescales 
of weeks to months), caused either by an instability in the mass
transfer from the secondary star (the MTI model) or by a thermal
instability in the accretion disc (the DI model) which switches
the disc from a low to a high viscosity regime (See, e.g., Pringle,
Verbunt \& Wade 1986).
On the other hand, nova-like variables seem to be permanently in a high
viscosity state, presumably as a result of the fact that the mass
transfer rate is always high.

The observation of accretion discs in CVs is well beyond the 
capabilities of current direct imaging techniques as it requires 
resolving structures on angular scales of micro arcseconds. 
However, the binary nature of these systems allows the application of
powerful indirect imaging techniques such as eclipse mapping (Horne 1985)
and Doppler tomography (Marsh \& Horne 1988) to probe the dynamics, 
structure and the time evolution of their accretion discs.

IP Pegasi is an intensively studied eclipsing dwarf nova with an 
orbital period of 3.8~hr.
Doppler tomography of emission lines revealed the presence of 
conspicuous spiral structures during outburst (Steeghs, Harlaftis \& 
Horne 1997; Harlaftis et~al. 1999), in support of hydrodynamical
disc simulations (Armitage \& Murray 1998, Stehle 1999).
Tidally induced spiral shocks are expected to appear in dwarf novae 
discs during outburst as the disc expands and its outer parts feel 
more effectively the gravitational attraction of the secondary star.
Eclipse mapping close to outburst maximum helped to constrain the
location and orientation of the spiral structures and to show that
the gas in the spiral shocks has sub-Keplerian velocities 
(Baptista, Harlaftis \& Steeghs 2000a; hereafter BHS).

In this paper we present spectrally-resolved eclipse maps of the 
accretion disc of IP~Pegasi on the late decline from an outburst. 
The observations and the data reduction are described in 
section~\ref{data}. The details of the data analysis are given
in section~\ref{analysis}. In section~\ref{results} we investigate 
the disc structure at different wavelengths, we present and discuss
spatially-resolved disc spectra and the spectrum of the uneclipsed
light as well as the radial distribution of the disc brightness 
temperatures and of the H$\alpha$ line intensity.
The results are summarized in section~\ref{fim}.

\section{Observations} \label{data}

Time-resolved optical spectro-photometry ($\Delta\lambda= 5500-6800$ \AA,
dispersion of 2.3 \AA\ per pixel) covering one eclipse of IP~Peg was 
obtained with the 1.3-m Michigan-Dartmouth-MIT (MDM) telescope  
during the late decline from the 1993 May outburst.
The run started on May 26 at 9:41 UT and consists of 25 short exposures 
($\Delta t= 120\,s$) at a time resolution of 147\,s.
The observations were taken under good sky conditions but at relatively
high air masses ($X \simeq 2$) and cover a narrow phase range around
eclipse (from $-0.1$ to $+0.15$ cycle).

The data were bias-subtracted and corrected for flat-field effects, and 
the 1-D spectra were optimally extracted with the algorithm of Horne 
(1986). Arc-lamp observations were used to calibrate the wavelength scale.
Observations of the standard spectrophotometric stars Feige 67 and 
BD+26\,2606 (Oke \& Gunn 1983; Massey et~al. 1988) were used to derive 
the instrumental sensitivity function and to flux calibrate the set of 
extracted spectra to an accuracy of better than 5 per cent. 
Error bars were computed taking into account the photon count noise 
and the sensitivity response of the instrument.
We corrected for slit losses by comparing simultaneous $V$ band photometry
provided by Zhang (1993, private communication) with synthetic photometry
computed from our spectra. However, since the data were acquired at an 
airmass of up to 2.5 there is a danger that the red end of the calibrated
spectra are contaminated by color dependent slit losses due to differential 
atmospheric refraction (estimated uncertainty of less than 20 per cent).

\begin{figure}
\includegraphics[bb=2cm 2.7cm 20cm 24cm,angle=-90,scale=0.38]{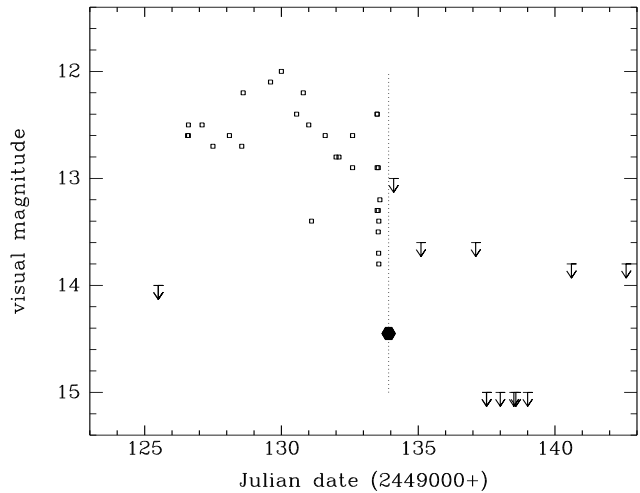}
\caption{ Visual light curve of IP~Peg for the epoch May--June 1993, 
 constructed from observations made by the AAVSO (squares). Arrows 
 indicate upper limits on the visual magnitude. A vertical dotted line
 marks the epoch of our observations. The synthetic $V$ band out-of-eclipse
 magnitude from our dataset is shown as a filled hexagon. }
\label{fig1}
\end{figure}

Figure \ref{fig1} shows the visual light curve of IP~Peg for the period
May--June 1993 from AAVSO observations. A vertical dotted line marks the 
epoch of our observations and a filled hexagon shows the corresponding 
out-of-eclipse synthetic $V$ band magnitude. 
%
%
%
%
The outburst started roughly on JD 2449126.
Our observations were performed $\simeq 8$ days after the onset of the
outburst and capture the late stages of the fast decline from maximum, when 
the star was about 0.5 mag above its quiescent (optical) brightness level.
HST observations on the following nights show that in the ultraviolet 
IP~Peg was still declining in brightness 3 days after our observations 
(Baptista et~al. 1994a).

\begin{figure}
\includegraphics[bb=2cm 3.2cm 19.5cm 24cm,angle=-90,scale=0.39]{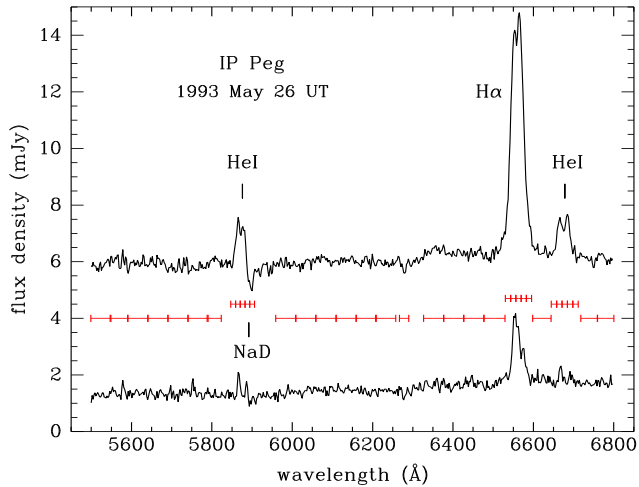}
\caption{Out-of-eclipse (phase range $+0.1$ to $+0.14$ cycle) and 
   mid-eclipse (phase range $-0.024$ to 0.0 cycle) spectra of IP~Peg 
   during the 1993 May outburst. These observations correspond to the 
   eclipse cycle 22240 (Wolf et~al. 1993). Horizontal ticks mark the 36
   narrow passbands used to extract light curves. }
\label{fig2}
\end{figure}

Average out-of-eclipse and mid-eclipse spectra are shown in
Fig.\ \ref{fig2}. The out of eclipse spectrum is dominated by strong,
broad and double-peaked H\,I and He\,I emission lines, as previously 
seen in outburst spectra (Pich\'e \& Szkody 1989; Marsh \& Horne 
1990). The interstellar Na\,D line is also visible.
The H$\alpha$ line has an equivalent width of $EW=(48 \pm 1)$~\AA.
There is still significant H$\alpha$ emission during eclipse whereas
the strength of the He\,I lines is considerably reduced at these phases
(the net line emission flux reduces by factors of about 5 and 15, 
respectively for H$\alpha$ and the He\,I lines),  
indicating that the He\,I lines comes from closer to the disc centre 
than H$\alpha$.
%
%
%
%

\section{Data Analysis} \label{analysis}

\subsection{Light curves}

The spectra were divided into a set of 21 narrow continuum passbands 
($\sim 50$ \AA\ wide) and 15 velocity-resolved passbands across the
H$\alpha$, He\,I $\lambda 5876$ and He\,I $\lambda 6678$ emission lines
(resolution of $600\; km\, s^{-1}$ per bin), in a total of 36 passbands
(indicated in Fig.~\ref{fig2}). 
In order to define the velocity-resolved passbands, a systemic velocity of
$\gamma= +50\; km\,s^{-1}$ (Marsh 1988; Hessman 1989) was subtracted from
the rest wavelength of each pixel; the spectral bins were defined by central 
velocity and velocity shift, and the average flux per bin was computed by
summing across the bin with partial pixel arithmetic (each pixel was 
weighted by the fraction of its width that is included in the passband).
Errors in the average flux per bin were also computed including partial 
pixel arithmetic.
For those passbands including emission lines the light curve comprises 
the total flux at the corresponding bin with no subtraction of a possible
continuum contribution 
\footnote{This choice allows one to simultaneously probe regions in which 
the lines are in absorption or emission with a single eclipse map by simply 
subtracting an average adjacent continuum map from the total flux line map
to derive a net emission/absorption line map.}.

Light curves were constructed for each passband by computing the average 
flux on the corresponding wavelength range and phase folding the results
according to the ephemeris of Wolf et~al. (1993),
\begin{equation}
T_{\rm mid}({\rm HJD}) = 2\,445\,615.4156 + 0.158\,206\,16 \; E \; ,
\label{efem}
\end{equation}
where $T_{mid}$ gives the inferior conjunction of the white dwarf.
IP~Peg shows long-term ($\simeq 5$ years) cyclical orbital period changes, 
with departures of the observed mid-eclipse timings of up to 2 minutes 
with respect to the ephemeris of Eq.(\ref{efem}). 
In order to correctly phase our light curves, we measured white dwarf 
egress times from quiescent data a few months before and after this 
outburst (Baptista et~al. 1994a). 
From these timings we inferred a white dwarf mid-eclipse phase of 
$\phi_0=-0.0082$ cycle for this epoch (adopting an eclipse width of
$\Delta\phi= 0.0863$ cycles, Wood \& Crawford 1986) and corrected the 
data accordingly in order to make the centre of the white dwarf eclipse
coincident with phase zero. 

In the standard eclipse mapping method all variations in the eclipse light
curve are interpreted as being caused by the changing occultation of the
emitting region by the secondary star.
Thus, out-of-eclipse brightness changes (e.g., orbital modulation due to
anisotropic emission from the bright spot) have to be removed before the
light curves can be analyzed.
The small differences in the out-of-eclipse brightness level before and
after eclipse (of up to 10 per cent) were removed by fitting a straight
line to the phases outside eclipse, dividing the light curve by the fitted
line, and scaling the normalized light curve to the value at phase zero. 
This procedure removes orbital variations outside eclipse with only minor 
effects on the eclipse shape itself.

\subsection{Eclipse Maps} \label{mem}

The light curves were analyzed with eclipse mapping techniques 
(Horne 1985; Baptista \& Steiner 1993) to solve for a map of the disk 
brightness distribution and for the flux of an additional uneclipsed 
component in each band. The reader is referred to Baptista (2001) for 
a recent review on the eclipse mapping method.

For our eclipse maps we adopted a flat grid of $51 \times 51$ pixels
centered on the primary star with side 2~R$_{L1}$, where R$_{L1}$ 
is the distance from the disk center to the inner Lagrangian point.
The eclipse geometry is defined by the mass ratio $q$ and the 
inclination $i$. The mass ratio $q$ defines the shape and the relative 
size of the Roche lobes. The inclination $i$ determines the shape and 
extension of the shadow of the secondary star as projected onto the 
orbital plane.
We adopted the values derived by Wood \& Crawford (1986), $q=0.5$ and
$i=81\degr$, which corresponds to an eclipse phase width of 
$\Delta\phi= 0.0863$ cycles. This combination of parameters ensures 
that the white dwarf is at the centre of the map.
Eclipse maps obtained using a different set of parameters, $q=0.58$ and 
$i=79.5\degr$ (Marsh 1988) show no perceptible difference with respect
to the maps derived using the above geometry. Hence, for the remainder 
of the paper we will refer to and show the results for ($q=0.5 \;, \;
i=81\degr$).

For the reconstructions we adopted the default of limited azimuthal 
smearing of Rutten, van Paradijs \& Tinbergen (1992a), which is better 
suited for recovering asymmetric structures than the original default 
of full azimuthal smearing (cf. Baptista, Steiner \& Horne 1996).
The reader is referred to BHS and Baptista (2001) for eclipse mapping
simulations with asymmetric sources which show how the presence of 
spiral structures affect the shape of the eclipse light curve, and 
which evaluate the ability of the eclipse mapping method to reconstruct 
asymmetric structures in eclipse maps.

The statistical uncertainties of the eclipse maps were estimated with a
Monte Carlo procedure (e.g., Rutten et al. 1992a). 
For a given narrow-band light curve a set of 20 artificial light curves 
is generated, in which the data points are independently and randomly 
varied according to a Gaussian distribution with standard deviation equal 
to the uncertainty at that point. The light curves are fitted with the 
eclipse mapping algorithm to produce a set of randomized eclipse maps. 
These are combined to produce an average map and a map of the residuals 
with respect to the average, which yields the statistical uncertainty at 
each pixel.
The uncertainties obtained with this procedure are used to estimate the 
errors in the derived radial brightness and temperature distributions as
well as in the spatially-resolved spectra.

Light curves and greyscale plots of the corresponding eclipse maps are
shown in Figs.\ \ref{fig3}-\ref{fig5}. These will be discussed in detail 
in section~\ref{results}.

\section{Results}  \label{results}

\subsection{Disc structure} \label{struc}

In this section we compare eclipse maps at selected passbands in order
to study the structure of the accretion disc at different wavelengths.

\begin{figure*}
\includegraphics[bb=3.5cm 3.7cm 18cm 24cm,angle=-90,scale=0.77]{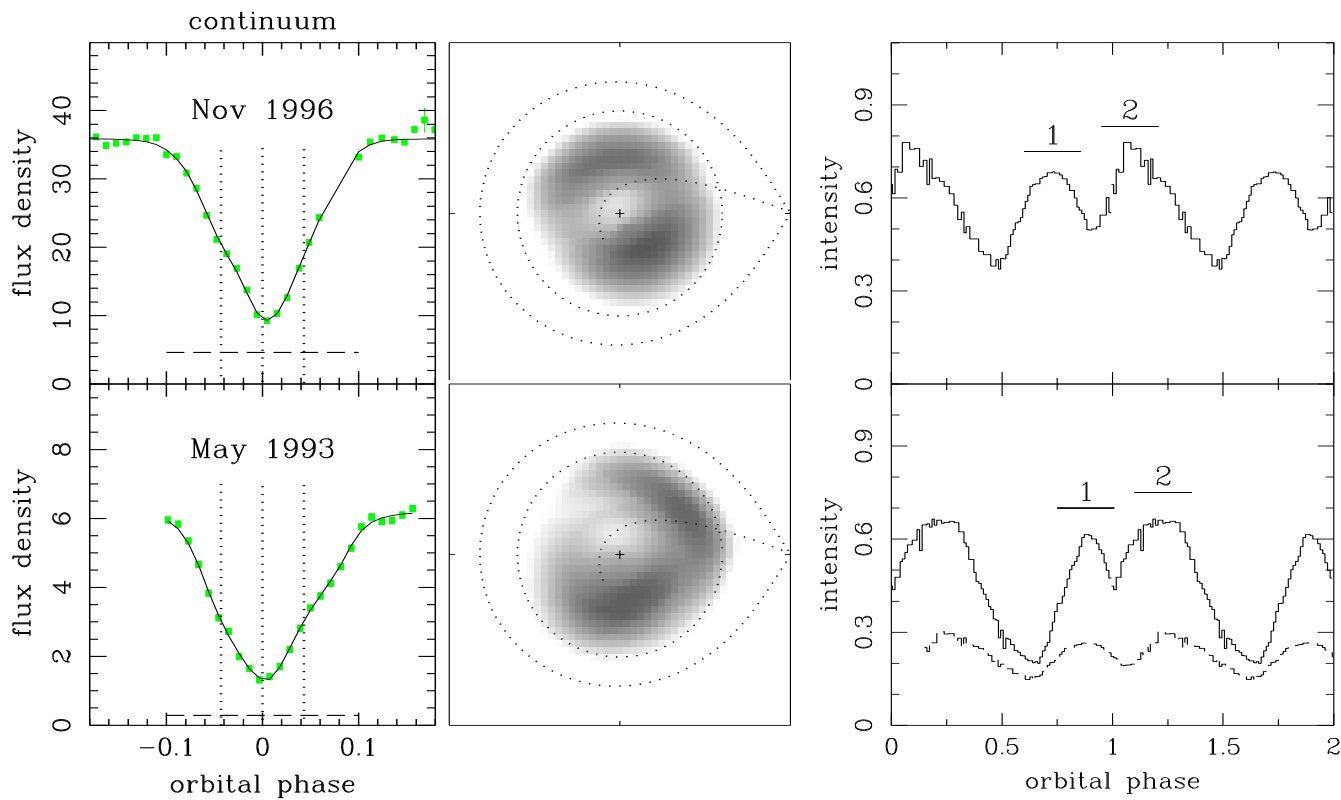}
 \caption{ Left: data (dots with error bars) and model (solid lines)
  light curves in the continuum for the November 1996 (BHS) and May 1993
  outbursts ($\lambda 6084$). Most of the error bars is smaller than 
  the plotted symbols. Vertical dotted lines mark mid-eclipse and 
  the ingress/egress times of the white dwarf. 
  Horizontal dashed lines indicate the uneclipsed component in each case. 
  Middle: the corresponding eclipse maps in a logarithmic greyscale. 
  Brighter regions are indicated in black; fainter regions in white.
  A cross marks the center of the disk; dotted lines show the Roche lobe,
  the gas stream trajectory, and a disc of radius $0.6\;R_{L1}$; the 
  secondary is to the right of each map and the stars rotate 
  counterclockwise. Right: The dependency of maximum intensity with binary
  phase for the maps in the middle panels. The intensities are plotted in
  an arbitrary scale. The location of the spiral arms are indicated by
  horizontal bars with labels 1 and 2. For comparison, the azimuthal
  distribution of the maximum intensity for the Nov 1996 map is plotted as
  a dashed line in the lower panel, displaced in phase by $+0.16$ cycle. }
 \label{fig3}
\end{figure*}

The lower panels of Fig.~\ref{fig3} show the light curve and the eclipse
map of the continuum passband at $\lambda 6084$.
%
%
%
%
The light curve exhibits an asymmetric `V' shape with broad bulges similar 
to those seen in the light curves of BHS and results in an eclipse map 
with two clear asymmetric arcs extended both in radius [$R\simeq 
(0.2-0.6)\; R_{L1}$] and in azimuth (by $\simeq 90\degr$), which we 
interpret as emission from spiral shocks -- as previously seen in 
Doppler tomograms (Steeghs et~al. 1997; Harlaftis et~al. 1999) and in
the eclipse maps of BHS. Similar eclipse shape and asymmetric structures
in the eclipse map are consistently found in all continuum passbands. 
Furthermore, since the eclipse in the simultaneous $V$ band light curve
shows the same `V' shape with broad bulges, we are confident that the
asymmetric structures in the eclipse maps are real and are not artifacts
caused, e.g., by an improper correction of the slit losses.
The continuum arcs contribute about 12 per cent of the total flux of
the eclipse map, which is comparable to the 13 per cent fractional 
contribution found by BHS despite the reduction by a factor of 6 in the 
out of eclipse flux level between the two outburst stages.
Hence, the spirals are still visible in the late stages of the outburst
and their fractional contribution to the continuum emission is similar 
to that measured close to outburst maximum.
This confirms and extends the results from Morales-Rueda, Marsh \& 
Billington (2000), who found that the spiral structure was still clearly 
visible in Doppler tomograms of IP~Peg 5 and 6 days after the start of 
an outburst, and also those of Steeghs et~al. (1996), the He\,I Doppler 
tomogram of which shows a clear spiral pattern some 8 days after the 
onset of the outburst.

Fig.~\ref{fig3} presents an illustrative comparison between our
results and those from BHS, which show the accretion disc of IP~Peg
close to the maximum of the 1996 November outburst.
Horizontal dashed lines in the left-hand panels indicate the uneclipsed
component in each case. While the uneclipsed component of the outburst
maximum map corresponds to about 12 per cent of the total flux (BHS),
it is reduced to about 5 per cent in our late decline map. If this light
arises from a vertically-extended disc wind (see below, and also section
\ref{fbg}) this comparison suggests that the wind emission (and possibly
the amount of ejected material) decreases faster than the disc brightness
at the late decline stage.
This is in line with previous suggestions by Pich\'e \& Szkody (1989) 
and Marsh \& Horne (1990), who attributed the low-velocity component seen
in He\,II $\lambda 4686$ to an outflowing wind from the inner disc regions
and pointed out that the reduction in strength of this feature as the 
outburst proceeds could be an evidence of a decreasing mass outflow in 
the wind.

The BHS continuum map corresponds to a shorter wavelength than our 
data ($\simeq \lambda 4400$). Nevertheless, since our experiment shows
that the continuum maps are reassuringly similar over a rather large 
spectral range ($\simeq 1200$ \AA), the following discussion should be
free from wavelength-dependent effects.

With respect to the arcs seen in the continuum map of BHS, the asymmetric
arcs in our continuum maps appear rotated in azimuth in the retrograde
sense (the disc gas and the secondary star in the maps of Fig.~\ref{fig3}
rotate counter-clockwise). 
In order to quantify this statement, we divided the eclipse maps in 
azimuthal slices (i.e., `slices of pizza') and computed the maximum 
intensity as well as the radius at which the intensity reaches a maximum 
for each azimuth. This exercise allows us to trace the distribution in 
radius and azimuth of the spiral structures. The azimuthal distribution 
of the maximum intensity for each map is plotted in the rightmost panels 
of Fig.\,\ref{fig3} as a function of orbital phase (orbital phases 
increases clockwise in the eclipse maps of Fig.\,\ref{fig3} and phase 
zero coincides with the inner lagrangian point L1). 
The location of the spiral arms are indicated by horizontal bars with 
labels 1 and 2. Assuming that the spiral structure which develops
is repeatable from one outburst to another, 
the difference in azimuth (orbital phase) of the spiral arms from one 
map to the other is clearly visible. By cross-correlating the two 
distributions we find a phase offset of $0.16\pm0.01$ cycle
(or, equivalently, a change in azimuth of $58\degr$).
For illustration purposes, the azimuthal distribution of the maximum 
intensity for the Nov 1996 map is plotted as a dashed line in the 
right-hand lower panel, displaced in phase by $+0.16$ cycle. 
It therefore seems that the orientation of the spiral arms has changed
from outburst maximum to the late decline. 

Changes in the orientation of the spiral arms of similar amplitude
($\Delta\theta \simeq 45-50 \degr$) are clearly seen in the sequence of 
He\,II $\lambda 4686$ tomograms of U~Gem on the decline from an outburst
(Groot 2001), whereas hints of the same effect in IP~Peg can be found 
by comparing the He\,I $\lambda 6678$ Doppler tomogram of Steeghs et~al. 
(1997) with that of Steeghs et~al. (1996).
Possible explanations include a thickness effect (changes in vertical 
thickness along the spiral arms leading to different apparent orientations), 
an optical depth effect (changes in intensity along the spiral pattern 
due to variable obscuration, also leading to different apparent 
orientations), or a geometrical effect (a real change in the azimuth of 
the spirals).

We may exclude the possibility that the asymmetric structures are
caused by density enhancements in an elliptical precessing disc
(instead of tidally induced spiral shocks) because in this case there 
would be an even probability of framing the structures (fixed in the
frame rotating with the precessing disc) at any orientation
-- which is hard to reconcile with the fact that in all Doppler 
tomograms and eclipse maps the asymmetries are seen roughly at the same 
azimuths.
  
\begin{figure*}
\includegraphics[bb=2cm 4.5cm 20cm 24cm,angle=-90,scale=0.6]{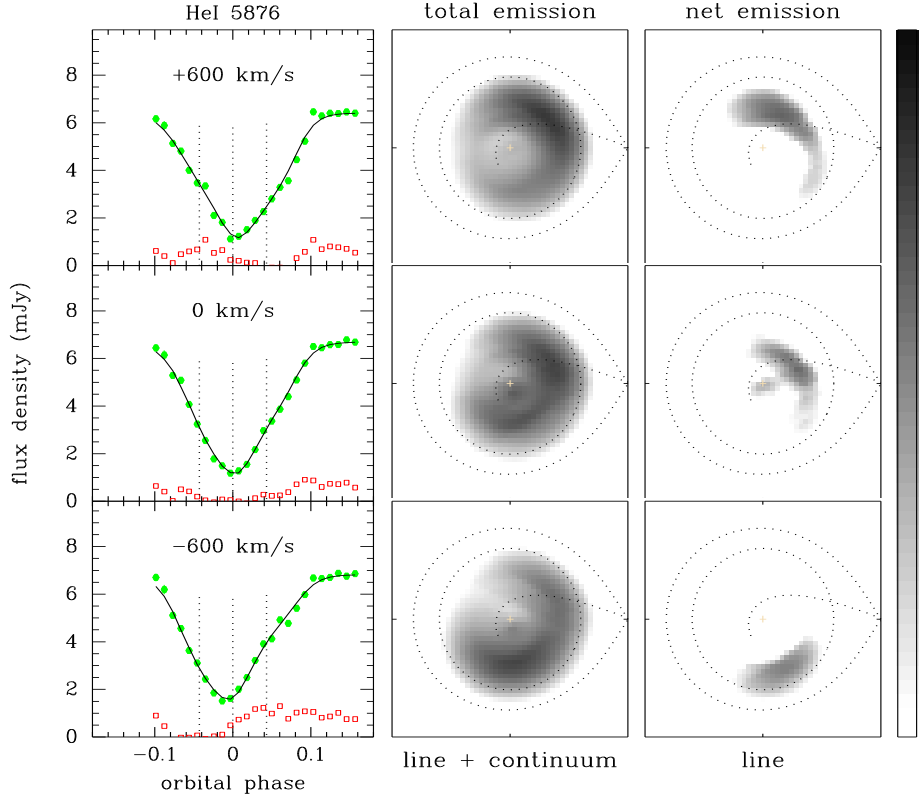}
 \caption{ Velocity-resolved light curves and eclipse maps in He\,I
  $\lambda 5876$. The open symbols depict the corresponding net emission 
  line light curve.
  The eclipse maps in the middle panel correspond to the total emission
  (line + continuum), while those in the right-hand panels show the net
  line emission. The notation is similar to that of Fig.~\ref{fig3}. }
 \label{fig4}
\end{figure*}

\begin{figure*}
\includegraphics[bb=2cm 4.5cm 20cm 24cm,angle=-90,scale=0.6]{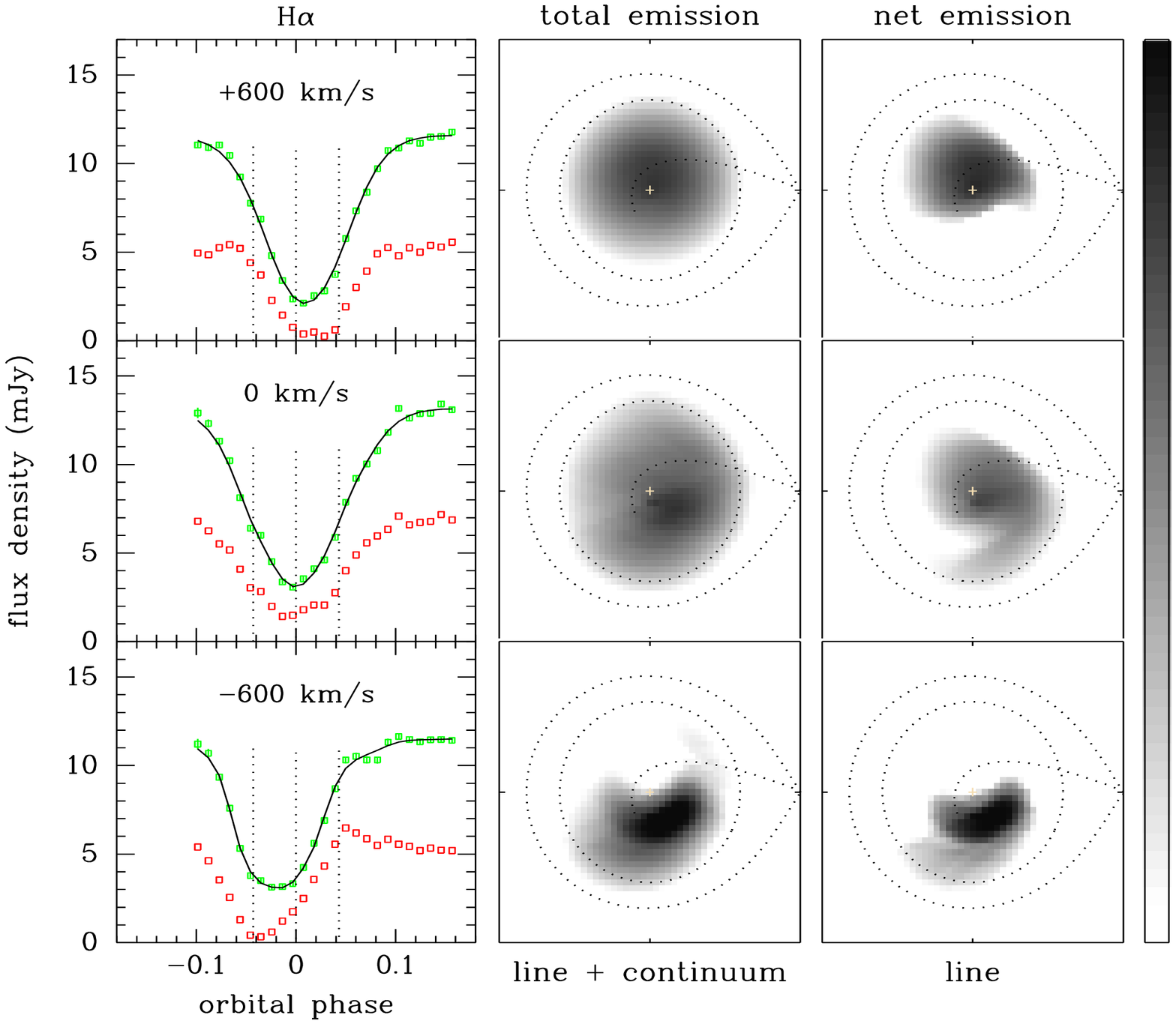}
 \caption{ Velocity-resolved light curves and eclipse maps in H$\alpha$.
 The notation is the same as in Fig. \ref{fig4}. }
 \label{fig5}
\end{figure*}

We now turn our attention to the disc structures in the emission
line maps. Velocity-resolved light curves and eclipse maps in He\,I
$\lambda 5876$ and H$\alpha$ are shown, respectively, in Figs.~\ref{fig4}
and \ref{fig5}. For illustration purposes, open symbols depict the
corresponding net emission line light curves (obtained by subtraction
of an average of the two adjacent continuum light curves in each case).
%
%
%
%
%
%
The eclipse maps in the middle panels correspond to the total emission 
(line + continuum), while those in the right-hand panels show the 
net line emission. The net emission maps were derived by combining the
continuum maps on both sides of the line and subtracting the resulting 
average continuum map from the line + continuum map. Only the regions 
in which the intensities are positive are shown in the net emission maps.

Hereafter, we will refer to the line + continuum light curve (or eclipse
map) simply as the line light curve (or eclipse map).

The He\,I line light curves display the same asymmetric `V' shape of the
continuum bands, but the strength of the bulges changes with velocity. 
The sequence of light curves across the line (Fig.~\ref{fig4},
left-hand panels) shows the expected behaviour for the eclipse of gas 
rotating in the prograde sense (the classical rotational disturbance), 
with the blue side of the line ($-600\; km \;s^{-1}$) being eclipsed 
earlier than the red side ($+600\; km \;s^{-1}$).
However, the eclipse maps in the symmetric velocity bins do not show 
the reflection symmetry (over the line joining both stars) expected for
line emission from a symmetric Keplerian disc around the white dwarf.
In fact, the He\,I eclipse maps (middle panels) show the same two 
asymmetric arcs of the continuum maps, with the interesting addition
that the receding arc (i.e., the one in the upper right quadrant of the 
map) is stronger in the red side and the approaching arc (lower left 
quadrant) is stronger in the blue side of the line. 
Aside from the expected contribution by low (projected) velocity 
material running roughly along the disc regions closest to the secondary
star, the line centre map shows a compact, low-velocity emitting region at 
disc centre, emphasizing the non-Keplerian distribution of the emitting gas. 
The two remaining velocity-resolved maps (at $\pm 1200\; km \;s^{-1}$,
not shown in Fig.~\ref{fig4}) are hardly distinguishable from the 
neighbouring continuum maps, indicating that there is no significant 
He\,I (net) emission at high velocities.
Similar results are found for the He\,I $\lambda 6678$ line.
These results suggest that the high-velocity emitting regions close 
to the white dwarf are veiled by low-velocity material,
in agreement with the fact the He\,I line profiles are relatively 
narrow (both lines disappear in the continuum at velocities of about 
$\pm 1200\; km \;s^{-1}$).
This could be another manifestation of outflowing gas in a disc
wind arising from close to the white dwarf.

Thus, the double-peaked profiles of the He\,I lines are not from 
a symmetric Keplerian disc but are produced by the two-armed spirals. 
Since the spirals extend up to large disc radii, the outer parts of 
them are still visible at mid-eclipse, leading to the residual He\,I
double-peaked emission seen in the mid-eclipse spectrum (Fig.\,\ref{fig2}).

A rotational disturbance effect is also clearly visible in the 
H$\alpha$ light curves. This leads to eclipse maps in which the 
intensity distribution moves from the approaching to the receding side
of the disc for increasing velocities.
There are, however, significant differences with respect to the He\,I 
line. The H$\alpha$ eclipse shapes are narrower and more symmetric than 
those of He\,I, leading to single-mode brightness distributions
concentrated towards the inner disc regions. As with the He\,I maps, 
there is no reflection symmetry for the eclipse maps of the symmetric 
velocity bins, while the line centre map reveals that most of the 
inner disc is covered by low velocity gas -- again indicating that a 
substantial contribution to the emission comes from sources other 
than gas in Keplerian orbits.
Perhaps more important, there is no evidence of the two-armed
spiral structure clearly seen in the continuum and He\,I maps
(the strong asymmetric structure seen in the blue bin map sits at a 
different azimuth and at a much smaller radius than the spiral arm in
the continuum and He\,I maps).
This suggests that the H$\alpha$ emission arises from a large, 
vertically-extended and optically thick region that veils the 
high-velocity gas close to disc centre and the continuum emission 
from the spiral shocks. We interpret this region as an 
outflowing and opaque (close to the orbital plane) disc wind.
Further evidence in support of this interpretation will be given
in section~\ref{fbg}.

Steeghs et~al. (1996) found a narrow ($\sim 150\; km\;s^{-1}$), nearly
stationary H$\alpha$ emission component in IP~Peg at a similar outburst 
stage (some 8 days after the onset of the outburst) which is only 
partially occulted during eclipse. They attributed this emission to 
stationary slingshot prominences caused by extended magnetic loops from 
the chromospherically active secondary star. 
Unfortunately, our relatively wide ($600\; km\;s^{-1}$) line centre 
passband is not appropriate to map/resolve such narrow, low-velocity 
component.

We end this section by remarking that 
conservation of angular momentum implies that any extended emitting
region around the white dwarf should display the classical rotational
disturbance. However, our results reveal that neither the He\,I nor
the H$\alpha$ lines are produced in a Keplerian disc.
The He\,I lines trace the two armed spiral shocks, while the H$\alpha$
line probably traces an outflowing (and spiraling) disc wind.
These results are consistent with the Doppler maps of Steeghs et~al.
(1996), who show that 8 days after the onset of the outburst the
spirals are still clear in He\,I but are no longer evident in H$\alpha$.

\subsection{Spatially-Resolved Spectra} \label{spec}

Each of the eclipse maps yield spatially-resolved information about 
the emitting region on a specific wavelength range.
By combining all narrow-band eclipse maps we are able to isolate 
the spectrum of the accretion disc for any desired region.

\begin{figure}
\includegraphics[bb=0.7cm 2.6cm 20.7cm 20cm,angle=-90,scale=0.4]{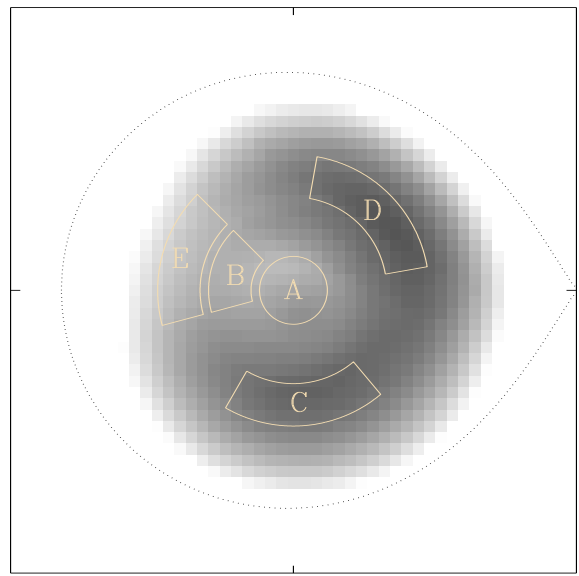}
 \caption{ Annular regions (A-F) used to extract spatially resolved spectra.
  A dotted line shows the projection of the primary Roche lobe onto the 
  orbital plane. }
 \label{fig6}
\end{figure}

The annular regions defined for the extraction of spatially-resolved 
spectra are shown in Fig.~\ref{fig6}.
In order to investigate the emission from the different light sources
we divided the maps into five annular sections aimed to separate the 
emission from the inner disc regions (labeled as regions A and B) and 
the spiral arms (regions C and D) as well as a spiral-free disc 
region at the same distance from disc centre (region E).
%
%
%
%
The central annular region, A, has a radial width of $\Delta R=0.12\; 
R_{L1}$ while the remaining regions have widths of $\Delta R=0.15\; 
R_{L1}$. Regions C, D and E cover the same range of radii, namely, 
$R= (0.33-0.48) \;R_{L1}$. 
The range in azimuth covered by the regions B and E was chosen in order 
to minimize contamination by light from the spiral arms.
Each spectrum is obtained by averaging the intensity of all pixels 
inside the corresponding annulus and the statistical uncertainties
affecting the average intensities are estimated with the Monte Carlo
procedure described in section\,\ref{mem}.
The spectra are given in units of flux density per pixel which is,
therefore, independent of the assumption about the distance to IP~Peg.
Changes in the definition of the annular regions by $0.05\;R_{L1}$ in 
radius and radial width and by $15 \degr$ in azimuth and azimuthal range
result in changes in the extracted spectra of the same order of the
quoted statistical uncertainties.

\begin{figure}
\includegraphics[bb=1.5cm 2.2cm 20cm 26.8cm,scale=0.47]{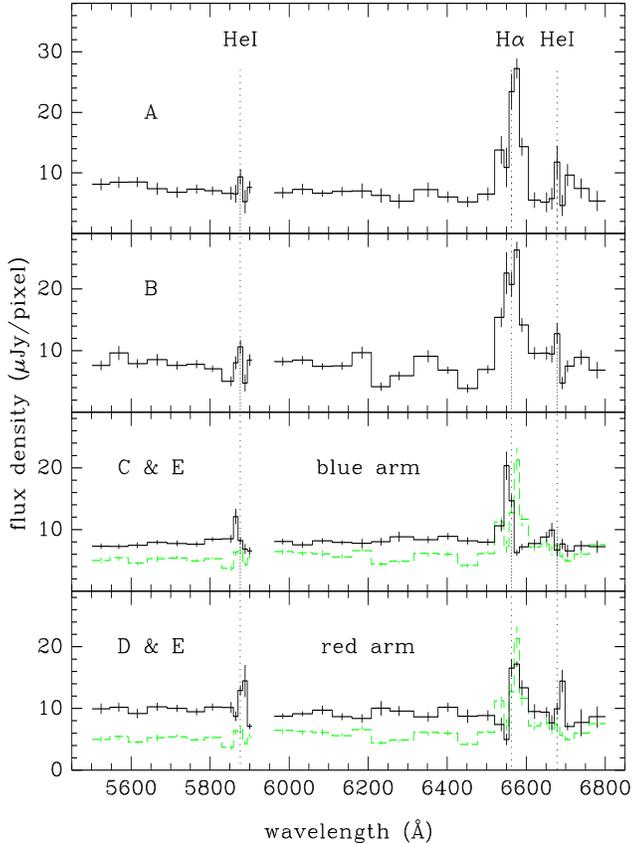}
 \caption{ Spatially-resolved spectra, obtained by computing the average
 intensity of each map in the regions defined in Fig.~\ref{fig6}.
 The uncertainties (vertical bars) were obtained by Monte Carlo
  simulations with the corresponding light curves. The spectrum of 
  region E is shown as a grey dashed line in the two lower panels. }
 \label{fig7}
\end{figure}

The spatially-resolved spectra are displayed in Fig.~\ref{fig7}.
The inner disc regions show an emission line spectrum with a strong
and broad H$\alpha$ component superimposed on a flat continuum.
This is in marked contrast with the results obtained for the nova-like
variables UX~UMa (Rutten et~al. 1994; Baptista et~al. 1998) and UU~Aqr
(Baptista et~al. 2000b), the inner disc regions of which show deep
and narrow absorption lines over a strong blue continuum.
Since the binary inclination of IP~Peg is similar to that of UU~Aqr 
($i=78\degr$, Baptista, Steiner \& Cieslinski 1994b), this difference 
cannot be attributed to inclination effects. The orbital periods and
mass ratios of these two binaries are also comparable (UU~Aqr has 
$P_{orb}= 3.9\; hr$ and $q=(0.3-0.4)$, Diaz \& Steiner 1991, 
Baptista et~al. 1994b).
Hence, this result suggests that different physical conditions hold
in the inner disc regions of these binaries.
%
%
%
%

The differences between the H$\alpha$ and He\,I lines are clear.
The H$\alpha$ line has a $FWZI\simeq 3000\; km\;s^{-1}$ in the inner
disc regions (A and B) which decreases to $FWZI\simeq 1800\; km\;s^{-1}$
in regions C, D and E, whereas the He\,I emission lines are narrow 
everywhere, with a $FWZI\simeq 600\; km\;s^{-1}$.
The greatest contribution to the integrated H$\alpha$ emission comes 
from the inner disc regions. The equivalent width of H$\alpha$ is 
$EW= (100\pm 2)$~\AA\ in regions A and B (twice that of the integrated, 
out of eclipse spectrum), 
and reduces to $EW\simeq 20$~\AA\ in the outer regions C, D and E.
The strength of the H$\alpha$ line is the same in the spirals and in
the spiral-free disc region E, emphasizing that H$\alpha$ is not a 
good tracer of the spiral arms at this late outburst stage.
On the other hand, the He\,I lines are dominated by contributions from
the spiral arms. The equivalent width of the He\,I lines at regions C 
and D is $EW= 5-10$~\AA, while they are weakly in emission in the 
inner disc regions ($EW\simeq 1$~\AA) and disappear in the continuum in 
the outer disc (region E).
This underscores the conclusion of section~\ref{struc} that the 
H$\alpha$ and the He\,I lines are produced in different regions.

All lines are seen blueshifted in the spectrum of the approaching arm
(labeled blue arm in Fig.~\ref{fig7}) and redshifted in the spectrum
of the receding arm (red arm), which is another way of seeing that
the gas rotates in the prograde sense.
The comparison of the spectrum of region E with those of regions C and D
shows that the continuum in the spirals is slightly bluer than in the
spiral-free disc region at same radii, suggesting that the gas in the
spirals is somewhat hotter than in the neighbouring disc regions.
The average continuum flux level is roughly the same in the inner and
outer, spiral-free disc regions, $f_\nu\simeq (6-8)\; \mu\!Jy\; pixel^{-1}$,
indicating a flat radial temperature distribution in the accretion disc
(section~\ref{radtemp}).

\subsection{The uneclipsed light} \label{fbg}

The uneclipsed component was introduced in the eclipse mapping method
to account for the fraction of the total light which is not coming from 
the accretion disc plane (e.g., light from the secondary star or from
a vertically-extended disc wind).
However, detailed simulations by Wood (1994) show that it is usually 
impossible to distinguish between a flared disc and an uneclipsed component 
to the total light. Both effects lead to the appearance of spurious 
structures in the disc regions farther away from the secondary star
(the left hemisphere of the eclipse maps of Figs. \ref{fig3}, \ref{fig4}
and \ref{fig5}, hereafter called the `back' side of the disc) and 
eclipse maps obtained with either model may lead to equally good fits
to the light curve. 
Baptista \& Catal\'an (2001) pointed out that if the uneclipsed component 
is caused by an optically-thin, vertically-extended disc wind, 
the uneclipsed spectrum should show a Balmer jump in emission plus strong 
emission lines, while in the case of a flared disc the spurious 
uneclipsed spectrum should reflect the difference between the disc 
spectrum of the back (deeper atmospheric layers seen at lower effective
inclinations) and the front (upper atmospheric layers seen at grazing 
incidence) sides and should mainly consist of continuum emission filled 
with absorption lines.
Our spectral mapping results allow to test their prediction.

\begin{figure}
\includegraphics[bb=2cm 2.5cm 19.5cm 24cm,angle=-90,scale=0.38]{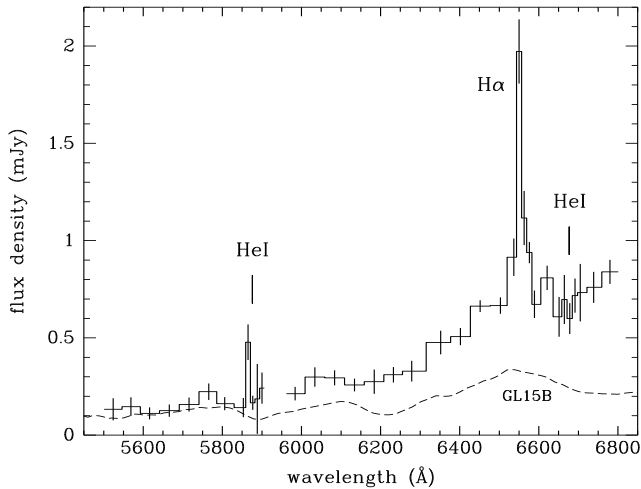}
 \caption{ The spectrum of the uneclipsed component. The dashed line
 is the spectrum of an M5V star (GL15\,B) scaled to a distance of 130\,pc. }
 \label{fig8}
\end{figure}

Fig.~\ref{fig8} shows the spectrum of the uneclipsed component.
It is dominated by a strong, blueshifted and narrow H$\alpha$ emission 
line superimposed on a red continuum. The He\,I $\lambda 5876$ line is 
marginally in emission while He\,I $\lambda 6678$ is absent. 
From these spectral features and the above discussion we conclude that
the observed uneclipsed component is not an artifact caused by erroneously
using a flat eclipse map to reconstruct a flared disc, but indeed 
corresponds to light coming from sources other than the accretion disc
in the orbital plane. 
%

In order to understand the uneclipsed spectrum, we first have to 
evaluate the contribution from the secondary star.
Martin, Jones \& Smith (1987) estimated the spectral type of the 
IP~Peg secondary star to be M3-5, while Catal\'an, Smith \& Jones (2001) 
used the TiO line-depth ratio to refine this estimate to M$(5.0 \pm 0.5)$V.
The former authors found a lower limit distance estimate of $130\,pc$
from the remaining infrared flux at mid-eclipse.
The spectrum of an M5V star scaled to a distance of $130\,pc$ is
depicted in Fig.~\ref{fig8} as a dashed line.
The star flux is reduced if the distance is made larger.
This exercise shows that, although the uneclipsed spectrum can easily
accommodate the light from an M5V secondary star, the emission from such
a star is not enough to account for the uneclipsed flux at the longer
wavelengths. Choosing a secondary star of later spectral type does not
overcome the mismatch (later spectral types lead to TiO absorption 
bands that are significantly stronger than observed). 
On the other hand, heating of the secondary star by irradiation from 
the hot, inner disc regions cannot be responsible for the extra uneclipsed 
light because the heated hemisphere (the one facing the white dwarf) 
is occulted during eclipse.
Thus, there must be another source of light contributing to the 
uneclipsed spectrum. 

The (extra) uneclipsed continuum increases towards longer wavelengths
suggesting that the Paschen jump is in emission. Together with the 
H$\alpha$ emission line, this indicates that the extra light consists 
of emission from optically thin gas from outside the orbital plane, 
probably arising in a vertically-extended disc wind. 
Our explanation for the inferences drawn from the H$\alpha$ eclipse 
maps (section~\ref{struc}) and the uneclipsed spectrum follows that 
of Baptista et~al. (2000b): the extended wind is dense and opaque close 
to the orbital plane (thus hiding the inner disc regions and the 
spirals from view) and becomes progressively transparent as the 
outflowing gas spreads over an increasing surface area (leading to
optically thin emission farther out of the orbital plane).

We estimate the fractional contribution of the uneclipsed, optically 
thin gas to the total flux by dividing the flux of the uneclipsed light 
(after subtraction of the contribution from an assumed M5V secondary 
star) by the average out of eclipse level at that passband.
The uneclipsed H$\alpha$ line flux amounts to 12 per cent of the total 
flux at that wavelength. The uneclipsed continuum contributes an 
increasing fraction of the total flux for longer wavelengths, reaching
10 per cent at the red end of the spectrum.
These numbers are significantly lower than the values of, respectively,
60 and 20 per cent found by Baptista et~al. (2000b) for the nova-like
UU~Aqr, probably reflecting the different physical conditions between 
the accretion discs of nova-like's and outbursting dwarf novae 
suggested in section~\ref{spec}. For example, the strong He\,I emission
lines in the uneclipsed spectrum of UU~Aqr (Fig.\,10 of Baptista et~al.
2000b) and the virtual absence of these lines in the spectrum of 
Fig.~\ref{fig8} suggests that the outflowing gas is cooler in IP~Peg.

Finally, we remark that the lack of evidence of a flared disc in the
uneclipsed spectrum, {\em per se}, is not enough to discard the 
possibility that the accretion disc of IP Peg is flared. If the half 
disc opening angle, $\alpha_d$, is so large that essentially all of 
the disc surface is hidden from view by the extended disc rim (as it 
should be if $\alpha_d \simgt 15\degr$), no front-back asymmetry 
is expected.

\subsection{Radial temperature distribution} \label{radtemp}

It has been a usual practice to convert the intensities in the eclipse 
maps to blackbody brightness temperatures and then compare them to the
radial run of the effective temperature predicted by steady state, 
optically thick disc models. A criticism about this procedure is that a
monochromatic blackbody brightness temperature may not always be a proper
estimate of the disc effective temperature. As pointed out by Baptista
et~al. (1998), a relation between these two quantities is non-trivial, 
and can only be properly obtained by constructing self-consistent models 
of the vertical structure of the disc. Nevertheless, the brightness 
temperature should be close to the effective temperature for the 
optically thick disc regions. 

Bobinger et~al. (1997) analyzed light curves of IP~Peg on the early 
decline from an outburst to find that the radial brightness temperature
distribution is essentially flat (with temperatures decreasing during
that outburst stage from about 9000 to $7000\;K$), in a clear
disagreement with the radial dependence predicted by the steady state
disk model ($T \propto R^{-3/4}$).
They raised various possibilities to explain the flatness of the 
temperature distribution (hole in inner disc regions, high disc rim,
optically thick wind sphere) without finding a compelling explanation
for it.

\begin{figure}
\includegraphics[bb=3cm 2.6cm 19.5cm 24cm,angle=-90,scale=0.38]{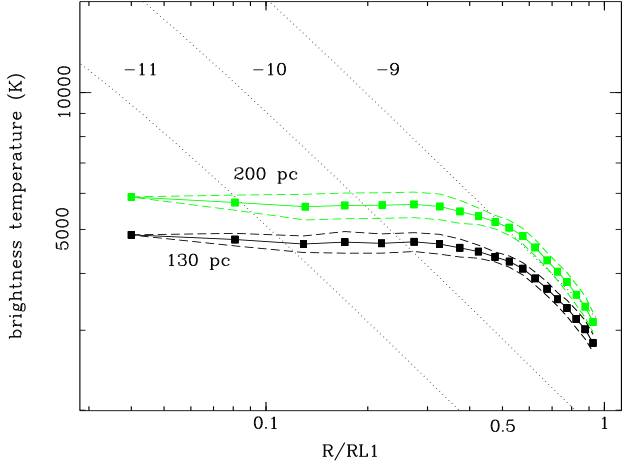}
 \caption{ The brightness temperature radial distribution, calculated
 assuming distances of 130\,pc (black curve) and 200\,pc (grey curve)
 to the system and no reddening. Dashed lines show the 1-$\sigma$ limit 
 on the average temperature for a given radius. Dotted lines correspond 
 to steady-state disc models for mass accretion rates of \.{M}$= 10^{-9}$,
 $10^{-10}$, and $10^{-11}\; M_\odot \; yr^{-1}$, assuming M$_1= 1.02 \;
 M_\odot$ and $R_1= 0.0075\;R_\odot$ (Marsh \& Horne 1990).
 The abscissa is in units of the distance from the disc centre to the
 inner Lagrangian point (R$_{\rm L1}$). }
 \label{fig9}
\end{figure}

A possibility not considered at that time is that the emission from 
the spiral arms distorts the results by artificially enhancing the 
intensities (and brightness temperatures) in the outer disc regions.
We tested this possibility by computing the $T(R)$ distribution for a 
disc section (slice of pizza) covering a range of azimuths identical to
those of regions B and E in order to minimize contamination by light 
from the spiral arms. 
The blackbody brightness temperature that reproduces the observed 
surface brightness at each pixel was calculated assuming distances of
130~pc and 200~pc to the system (with no reddening included). 
The disc section was then divided in radial bins of $0.05\;R_{L1}$ 
and a median brightness temperature was derived for each bin. 
These are shown in Fig.~\ref{fig9} as interconnected squares. 
%
%
The dashed lines show the 1-$\sigma$ limits on the average temperatures.
Steady-state disc models for mass accretion rates of $10^{-9}$, $10^{-10}$,
and $10^{-11}\; M_\odot$\,yr$^{-1}$ are plotted as dotted lines for 
comparison. 

The distribution is flat, in agreement with the results of Bobinger et~al.
(1997), with temperatures of about $5000\,K$ (for $D=130\,pc$) or $6000\;K$ 
(if $D=200\,pc$ is assumed) in the whole disc ($R \simlt 0.6\; R_{L1}$).
The temperatures are systematically lower than those derived by Bobinger
et~al. (1997), probably because our observations correspond to a later 
stage of the decline than theirs.
The agreement between the observed distribution and the $T\propto R^{-3/4}$
law in the outer regions of the map ($R\simgt 0.6\; R_{L1}$) is fortuitous:
it corresponds to the radii at which the intensities drop fast as the disc
rim is reached. 

We can therefore safely exclude the emission from the spiral arms as the 
cause of the observed flat radial temperature distribution of IP~Peg in
outburst.

As suggested by Morales-Rueda et~al. (2000) and Webb et~al. (1999), 
it could be that during outbursts the half disc opening angle in IP~Peg
is so large ($\alpha_d \simgt 15 \degr$) that essentially all of the 
disc surface remains hidden from view by a thick disc rim and only the
vertically-extended spiral arms and disc wind emission can be observed.
In this case the occultation of the thick disc rim by the secondary star
leads to a shallow, `V' shaped eclipse (if the rim has an approximately 
uniform surface brightness distribution) and the resulting eclipse map
naturally shows a flat (spurious) radial temperature distribution (see 
the simulations of Bobinger et~al. 1997).

However, the evidences in favour of a large disc opening angle in
IP~Peg during outbursts are not yet conclusive.
The large $\alpha_d$ values of the models by Webb et~al. (1999) are 
the direct consequence of the assumption that the outbursting disc 
brightness distribution is that of an optically thick steady-state disc
model\footnote{ 
	With the steep $T \propto R^{-3/4}$ distribution, the only way of 
	preventing an `U' shaped, sharp and deep eclipse is to make the disc 
	rim so high as to cover most (if not all) of the disc surface.
}.
Although this may indeed be the case during the plateau phase of
long-lasting outbursts, it is probably not generally true on the 
course of an outburst (see, e.g., Rutten et~al. 1992b and Baptista 
\& Catal\'an 2001). 
Doppler maps of the Balmer lines during outburst show that the line 
emission from the secondary star is displaced from the L1 point towards 
its centre of mass and that no emission is seen from its equator,
leading Morales-Rueda et~al. (2000) to the conclusion that the 
secondary star is shielded from the inner disc emission by a 
significantly thick disc rim ($H/R\simeq 0.2$).
These maps were computed assuming a semi-amplitude of the secondary
star radial velocity of $K_2= 300\; km\,s^{-1}$. The $K_2$ of IP~Peg
is uncertain by $\sim \pm\, 30\; km\,s^{-1}$ due to the effects of
irradiation (e.g., Warner 1995). If the revised value $K_2= 330\; 
km\,s^{-1}$ of Beekman et~al. (2000) is used, the Roche lobe of the
secondary star is displaced upward in the Doppler maps, therefore
pushing the line emission towards the L1 point and reducing the
need of a large disc opening angle to explain the observations.

The most promising interpretation put forward by Bobinger et~al. (1997)
to explain the flat radial temperature distribution and its time 
evolution on the decline from maximum is that of an optically thick 
and geometrically flat wind-layer over the disc surface decreasing in 
radius along the decline from outburst. 
This is consistent with the inferences drawn from the H$\alpha$ line 
maps (section~\ref{struc}).
However, this scenario also has problems. It predicts that any asymmetric
structure in the underlying disc would be screened by the opaque wind-layer,
which is hard to reconcile with the conspicuous spiral arcs seen in the
continuum and He\,I eclipse maps unless the vertical extent of the spiral 
arms is larger than that of the veiling wind-layer. On the other hand, the 
H$\alpha$ maps require that the photosphere of the opaque wind-layer in
this line is at a higher height than the spirals.

\subsection{H$\alpha$ surface brightness distribution}

\begin{figure}
\includegraphics[bb=2.8cm 2.7cm 18cm 25.3cm,scale=0.55]{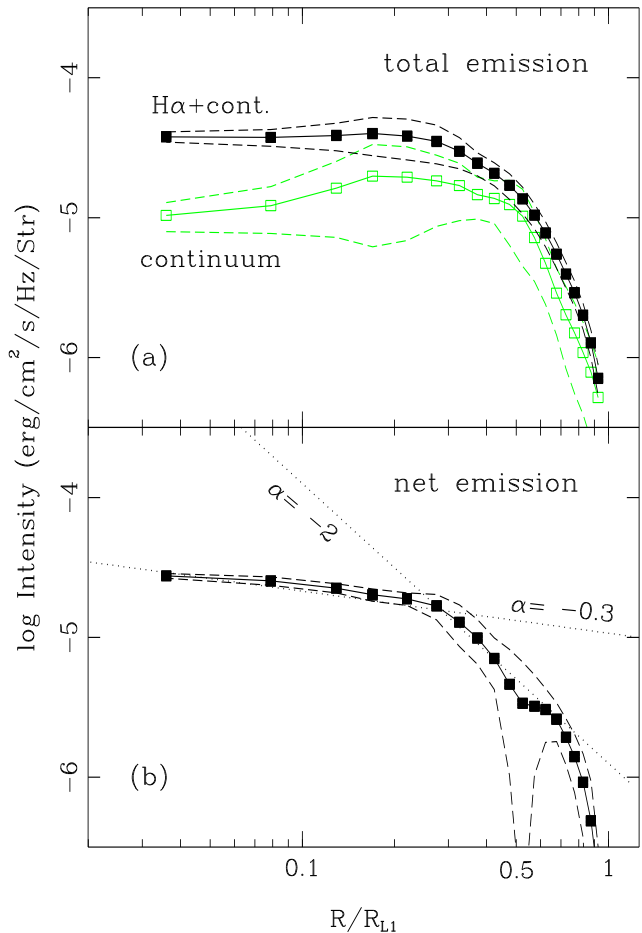}
 \caption{ (a) Radial intensity distributions for combined H$\alpha$
 and continuum (black) maps. The intensity scale assumes a distance of
 130\,pc. The notation is the same as in Fig.~\ref{fig9}. 
 (b) The H$\alpha$ net line emission radial distribution.
 Dotted lines depicts the slopes corresponding to $I_\nu(R) \propto
 R^{-0.3}$ and $I_\nu(R) \propto R^{-2}$ laws. }
 \label{fig10}
\end{figure}

The top panel of Fig.\,\ref{fig10} shows the radial intensity distribution 
for H$\alpha$ and the adjacent continuum in a logarithmic scale. 
%
%
The eclipse maps were divided in radial bins of $0.05\;R_{L1}$ 
and a median intensity was computed for each bin. 
These are shown in Fig.~\ref{fig10} as interconnected squares.
The dashed lines show the 1-$\sigma$ limits on the average intensity. 
The line distribution is obtained from the average of all
velocity-resolved maps across the line, while the continuum distribution
is obtained from the average of eclipse maps on both sides of the line. 
The H$\alpha$ net line emission surface brightness radial profile was
computed by subtracting the distribution of the adjacent continuum from 
that of the line, and is shown in the lower panel.
The large dispersion seen in the intermediate regions of the continuum
distribution reflect the large asymmetries caused by the spiral arms.
In the external map regions ($R \simgt 0.7\; R_{L1}$) the intensities
of both line and continuum drop by a factor $\sim 10^2$ with respect
to the inner disc regions, making the computation of the net emission
quite noisy and unreliable.

H$\alpha$ is seen in emission (intensities larger than those at the
adjacent continuum) at all disc radii and up to $R \simeq 0.7\; R_{\rm L1}$.
There are clearly two different emission regimes. 
The emission-line surface brightness is almost flat in the inner disc 
regions, $I \propto R^{-0.3}$ for $R<0.3\;R_{L1}$, but decreases sharply
with radius in the outer disc, $I\propto R^{-2}$ for $R>0.3\; R_{L1}$.
This is in line with the results of Marsh \& Horne (1990). They found
that in quiescence the Balmer lines show a steep surface brightness
distribution, $I\propto R^{-1.8}$, while in outburst the surface 
distribution flattens in the inner disc regions ($R\simlt 0.35\; R_{L1}$)
with a decreasing slope for the lower lines of the Balmer series.
The observed behaviour suggests that the emission from the outer disc regions 
may be powered by the same mechanism producing the quiescent Balmer lines, 
whereas the H$\alpha$ from the inner disc arises in the opaque outflowing 
wind by a different line emission mechanism, possibly photoionization by
radiation from the boundary layer (Marsh \& Horne 1990).

\section{Conclusions} \label{fim}

We analyzed time-resolved optical spectroscopy of the dwarf nova IP~Pegasi
on the late decline from the May 1993 outburst with eclipse mapping 
techniques in order to investigate the structure and the spectrum of its
accretion disc. The main results of this study can be summarized as follows:

\begin{enumerate}

\item The continuum light curves exhibit an asymmetric `V' shape with broad 
bulges and results in eclipse maps with two asymmetric arcs extended 
both in radius [$R\simeq (0.2-0.6)\; R_{L1}$] and in azimuth (by $\simeq
90\degr$), interpreted as a two-armed spiral shock. 
Our results reveal that the spirals are still visible in the late stages
of the outburst and their fractional contribution to the continuum 
emission is similar to that measured close to outburst maximum.

\item The asymmetric arcs in our continuum maps are rotated in azimuth by
$58\degr$ ($0.16\pm 0.01$ of the binary cycle) in the retrograde 
sense in comparison with the eclipse map of BHS.

\item Velocity-resolved light curves across the H$\alpha$ and the He\,I 
lines show the classical rotational disturbance, with the blue side of 
the line being eclipsed earlier than the red side. 
The spiral arms are clearly seen in the He\,I maps, with the receding 
arm being stronger in the red side whereas the approaching arm is 
stronger in the blue side of the line.
The double-peaked profiles of the He\,I lines are not from 
a symmetric Keplerian disc but are produced by the two-armed spirals. 
The analysis of the H$\alpha$ maps suggests that the H$\alpha$ emission
arises from a large, vertically-extended and optically thick region 
which we interpret as an outflowing (and spiraling) disc wind.

\item The inner disc regions show an emission line spectrum with a 
strong and broad H$\alpha$ component superimposed on a flat continuum.
This is in marked contrast with the results from the spectral mapping of
nova-like variables of comparable binary parameters and suggests that 
intrinsically different physical conditions hold in the inner disc 
regions of outbursting dwarf novae and nova-like systems.

\item The comparison of the spectrum of the spiral arms with that of
a spiral-free disc region at same radius confirms that the He\,I lines
are dominated by contribution from the spiral arms (with an
equivalent width of $EW= 5-10$~\AA) and suggests that the gas in the 
spirals is hotter than in the neighbouring disc regions.

\item The spectrum of the uneclipsed light is dominated by a strong, 
blueshifted and narrow H$\alpha$ emission line superimposed on a red 
continuum and can be understood as a combination of emission from an
M5V secondary star plus optically
thin emission from a vertically-extended disc wind. After subtraction
of the contribution of the secondary star light, the uneclipsed H$\alpha$
and continuum fluxes amount to, respectively, 12 and 10 per cent of the
total flux at the corresponding wavelengths.
These numbers are significantly lower than those found in spectral
mapping experiments of the nova-like variables UX~UMa and UU~Aqr and 
suggest that the wind emission in the late stages of the outburst in 
IP~Peg is considerably weaker than in those nova-like systems.

\item The previously observed flat radial temperature distribution of 
IP~Peg in outburst is not a distortion caused by the spiral arms. 
The radial temperature distribution computed from the spiral-free disc 
regions is still flat, with temperatures of about $5000\,K$ 
(for $D=130\,pc$) or $6000\;K$ (if $D=200\,pc$ is assumed) in the
whole disc ($R \simlt 0.6\; R_{L1}$).

\item H$\alpha$ is seen in emission (intensities larger than those at
the adjacent continuum) at all disc radii and up to $R \simeq 0.7\; 
R_{\rm L1}$. The emission-line surface brightness is flat in the inner 
disc regions, $I \propto R^{-0.3}$ for $R<0.3\;R_{L1}$, but decreases 
sharply with radius in the outer disc, $I\propto R^{-2}$ for 
$R>0.3\; R_{L1}$.

\end{enumerate}

\section*{Acknowledgments}

We thank an anonymous referee for valuable comments and suggestions on 
an earlier version of the manuscript.
In this research we have used, and acknowledge with thanks, data from
the AAVSO International Database and the VSNET that are based on
observations collected by variable star observers worldwide. We thank 
Er-Ho Zhang for kindly providing us with his unpublished $V$ band data.
This work was partially supported by the PRONEX/Brazil program through
the research grant FAURGS/CNPq 66.2088/1997-2.
RB acknowledges financial support from CNPq/Brazil through grant no. 300\,354/96-7.

\end{document}